\documentclass[12pt]{iopart}
\def\ket#1{|#1\rangle}
\def\bra#1{\langle#1|}

\usepackage{iopams}
\usepackage{graphicx}
\usepackage{cite}

\begin{document}

\title{Photonic quantum simulation of ground state configurations of Heisenberg square and checkerboard lattice spin systems}
\author{Xiao-song Ma$^{1,2}$, Borivoje Daki{\' c}$^{3}$, Sebastian Kropatschek$^{1}$, William Naylor$^{1}$, Yang-hao Chan$^{4,5}$, Zhe-xuan Gong$^{4,5}$, Lu-ming Duan$^{4,5}$, Anton Zeilinger$^{1,2}$, Philip Walther$^{3}$}

\address{$^1$~Institute for Quantum Optics and Quantum Information (IQOQI), Austrian Academy of Sciences, Boltzmanngasse 3, A-1090 Vienna, Austria}

\address{$^2$~Vienna Centre for Quantum Science and Technology (VCQ), Faculty of Physics, University of Vienna, Boltzmanngasse 5, A-1090 Vienna, Austria}

\address{$^3$~Faculty of Physics, University of Vienna, Boltzmanngasse 5, A-1090 Vienna, Austria}

\address{$^4$~Department of Physics and MCTP, University of Michigan, Ann Arbor, MI 48109, USA}

\address{$^5$~Center for Quantum Information, IIIS, Tsinghua University, Beijing, China}

\ead{Xiaosong.Ma@univie.ac.at}

\date{\today}

\begin{abstract}
Photonic quantum simulators are promising candidates for providing insight into other small- to medium-sized quantum systems. The available photonic quantum technology is reaching the state where significant advantages arise for the quantum simulation of interesting questions in Heisenberg spin systems. Here we experimentally simulate such spin systems with single photons and linear optics. The effective Heisenberg-type interactions among individual single photons are realized by quantum interference at the tunable direction coupler followed by the measurement process. The effective interactions are characterized by comparing the entanglement dynamics
using pairwise concurrence of a four-photon quantum system. We further show that photonic quantum simulations of generalized Heisenberg interactions on a four-site square lattice and a six-site checkerboard lattice are in reach of current technology.
\end{abstract}

\maketitle

\section{Introduction}
More than a quarter of a century ago, Richard Feynman~\cite{Feynman1982,Feynman1986a} envisioned that a well-controlled quantum mechanical system can be used for the efficient simulation of other quantum systems and thus would be capable of calculating properties that are unfeasible for classical computers. Quantum simulation promises potential returns in understanding detailed quantum phenomenon of inaccessible quantum systems, from molecular structure to the behavior of high-temperature superconductors~\cite{Lloyd1996,Buluta2009}. Moreover, quantum simulations are conjectured to be less demanding than quantum computations by being less stringent on explicit gate operations or error correction~\cite{Buluta2009}.

Due to these reasons, quantum simulation has led to many theoretical proposals~\cite{Aspuru-Guzik2005,Trebst2006,Kassal2008,Verstraete2009,Buluta2009}. Recently, various quantum simulators based on different physical platforms are being constructed, such as atoms in optical lattices~\cite{Lewenstein2007,Bakr2009,Bakr2010,Trotzky2010,Sherson2010,Weitenberg2011}, trapped ions~\cite{Friedenauer2008,Gerritsma2010,Kim2010,Barreiro2011,Islam2011,Lanyon2011}, nuclear magnetic resonance systems~\cite{Peng2009,Du2010}, superconducting circuits~\cite{Neeley2009}, as well as single photons~\cite{Lu2009,Pachos2009,Lanyon2010,Broome2010,Peruzzo2010,Lavoie2010,Ma2011,Matthews2011,Sansoni2012}.
Motivated by the seminal work of Knill, Laflamme and Milburn~\cite{Knill2001} photons are proven to be a suitable system for efficient quantum computing and quantum simulation. Precise single-photon manipulations and tunable measurement-induced two-photon interactions are the essential ingredients for photonic analog quantum simulation and have been demonstrated. In addition to high-level quantum control, such photonic quantum simulators can produce exotic entangled states which are important for understanding the many-body dynamics in quantum chemistry and solid-state physics~\cite{Kassal2008, Verstraete2009, Ma2011}.

Photons barely interact with the environment, which makes them ideal for exploiting various quantum phenomena. But in contrast to ions or atoms in optical lattices where physical interactions can be easily implemented via Coulomb interactions or Feshbach resonances, photonic simulations require various preparation steps to achieve controllable effective interactions. First, instead of non-correlated particles, one creates entangled singlet pairs of photons from non-linear crystals and these are used as input states for quantum simulation. In the language of quantum information we would say that each photon brings half an ebit (one half of the singlet) of correlation. Second, the non-correlated photons are overlapped at a tuneable directional coupler and measured in different output modes to introduce effective non-linearities that create the targeted many-body correlation. This is an effective interaction among the photons (for detailed calculation see Ref.\ \cite{Ma2011}). Note that the recently proposed universal quantum computation based on coherent photon conversion provides an alternative avenue for achieving photon-photon interaction~\cite{Langford2011}.

\section{Analog quantum simulations with photons and linear optics}

In the case of the simulation of spin-1/2 particles the photon's polarization is ideally suited as horizontally-polarized states $\ket{H}$ and vertically-polarized states $\ket{V}$ representing for example, spin-up and spin-down states. Moreover, the ability to prepare symmetric polarization-entangled states $\ket{\Phi}^{\pm}_{1,2}=\frac{1}{\sqrt{2}}(\ket{H}_{1}\ket{H}_{2}\pm\ket{V}_{1}\ket{V}_{2})$, $\ket{\Psi}^{+}_{1,2}=\frac{1}{\sqrt{2}}(\ket{H}_{1}\ket{V}_{2}+\ket{V}_{1}\ket{H}_{2})$ and the anti-symmetric state $\ket{\Psi^{-}}_{1,2}=\frac{1}{\sqrt{2}}(\ket{H}_{1}\ket{V}_{2}-\ket{V}_{1}\ket{H}_{2})$ enables the establishment of states with bosonic and fermionic character~\cite{Mattle1996,Lavoie2010,Ma2011}. The latter shares the same quantum correlations as Heisenberg-interacting spins or so-called valence bond states~\cite{Anderson1987}.

The theoretical investigation of strongly-correlated spin systems has led to few exact theorems which in some cases are of importance for the quantum simulation of chemical and physical models. In the particular case of a nearest-neighbor antiferromagnetic Heisenberg-interacting spin system it was shown by Marshall~\cite{Marshall1955} that the ground state for N spins on a bipartite lattice has total spin zero ($S^2= 0$).  This theorem and its extension~\cite{Lieb1962} lead to the fact that the ground state must be built as a linear superposition of singlet spin states or valence bonds. This constraint, that forces the ground state's total spin to be zero, gives rise to various valence-bond configurations that are either localized or fluctuating as superposition of different singlet partitionings. Localized configurations are typically referred to valence bond solids and delocalized valence-bond states correspond to frustrated quantum spin liquids or resonating valence-bond states~\cite{Affleck1987, Balents2010}. Recently, the photonic quantum simulation of a four spin-1/2 square lattice as an archetype system~\cite{Ma2011} showed that quantum monogamy~\cite{Coffman2000a,Osborne2006} plays an important role in frustrated Heisenberg spin systems.

Analog quantum simulation uses the advantages of digital and adiabatic quantum simulators by utilizing continuously tunable quantum gates. In Ref.\ \cite{Ma2011}, we reported the analog quantum simulation of the ground state to probe Heisenberg-type interactions of a spin-1/2 tetramer. The variable measurement-induced interactions between photons are crucial for our experiments. In conjunction with single-photon manipulation and detection, it allows us to show the dynamics of ground-state energies and pair-wise quantum correlations of this tetramer. Here, we present a detailed experimental study of these photon-photon interactions. Depending on the interaction strength, frustration within the system emerges such that the ground state evolves from a localized to a delocalized valence-bond state. The quantum monogamy relation is unambiguously demonstrated via entanglement measures~\cite{Coffman2000a,Osborne2006}.

\begin{figure}
 \begin{center}
\includegraphics[width=0.9\textwidth,keepaspectratio]{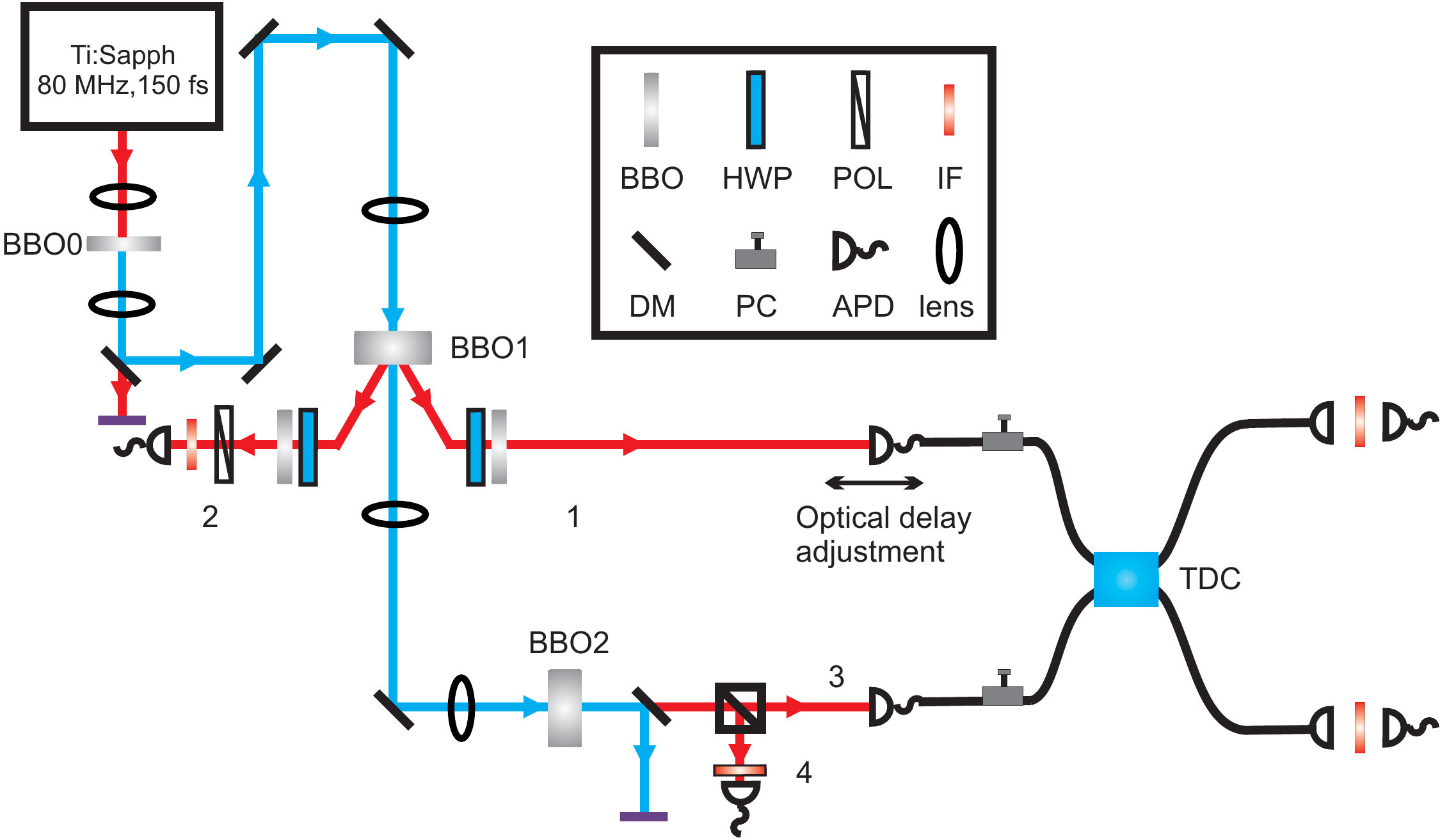}
\caption{\label{fig:setup}
Experimental setup for testing variable measurement-induced interaction between two independent photons. Ultraviolet (UV) femtosecond laser pulses are up-converted from a mode-locked Ti:sapphire laser. The laser pulses are up converted (BBO0) and the obtained temporal duration and repetition rate is 140~fs and~80 MHz, respectively. The UV pulses and the remaining fundamental pulses are separated with several dichroic mirrors (DM). Lenses are used to focus and collimate the UV pulses in order to achieve good up and down conversion efficiencies. One pair of the polarization entangled photons (photons 1 and 2) is generated from a $\beta$-barium borate crystal (BBO1) via spontaneous parametric down conversion (SPDC) in a non-collinear type-II phase matching configuration. Two half-wave plates (HWP) and compensating BBO crystals are used to counter walk-off effects in the down-conversion crystal. Another pair of correlated photons (photons 3 and 4) is generated from BBO2 in a collinear type-II phase matching configuration and are separated by a polarizing beam splitter (PBS). Single-mode fibers and interference filters (IF) are used to clean their spatial and spectral modes, respectively. We vary the path length difference between two photons with a motorized translation stage mounted on the fiber coupling stage of photon 1. Fiber polarization controllers (PC) are employed to eliminate the polarization distinguishability of the two interfering photons. The tunable measurement-induced interaction among photons is achieved by a tunable directional coupler (TDC), followed by a projective measurement of one photon in each of the four output modes.}
 \end{center}
\end{figure}

The experimental setup for studying variable measurement-induced interactions is shown in Fig.\ \ref{fig:setup}. Our pump source is a mode-locked Ti:sapphire femto-second laser with a pulse duration of 140~fs and a repetition rate of 80 MHz. The central wavelength of the pump is at 808 nm. Then we use a $\beta$-barium borate crystal (BBO0) to up-convert the pump pulses to ultraviolet (UV) pulses via second harmonic generation. The up-converted UV pulses' central wavelength is 404 nm with a power of 700 mW. Then we clean the UV pulses with several dichroic mirrors (DM).
Photons 1 and 2 are generated from a BBO crystal (BBO1) via spontaneous parametric down conversion (SPDC) in a non-collinear type-II phase matching configuration and in an polarization-entangled state after walk-off compensation~\cite{Kwiat1995}. Photons 3 and 4 are generated from BBO2 in a collinear type-II phase matching configuration and are separated by a polarizing beam splitter (PBS). We guide photons 1 and 3 to a tunable directional coupler (TDC) and then detect them by avalanche photodiodes (APD), which together enable the tunable measurement-induced photon-photon interactions to happen. The relative temporal delay between the photons is adjusted with a motorized translating stage mounted on the fiber coupler of photon 1. Fiber polarization controllers (PC) are employed to eliminate the polarization distinguishability of the two interfering photons.

\begin{figure}
 \begin{center}
\includegraphics[width=1\textwidth]{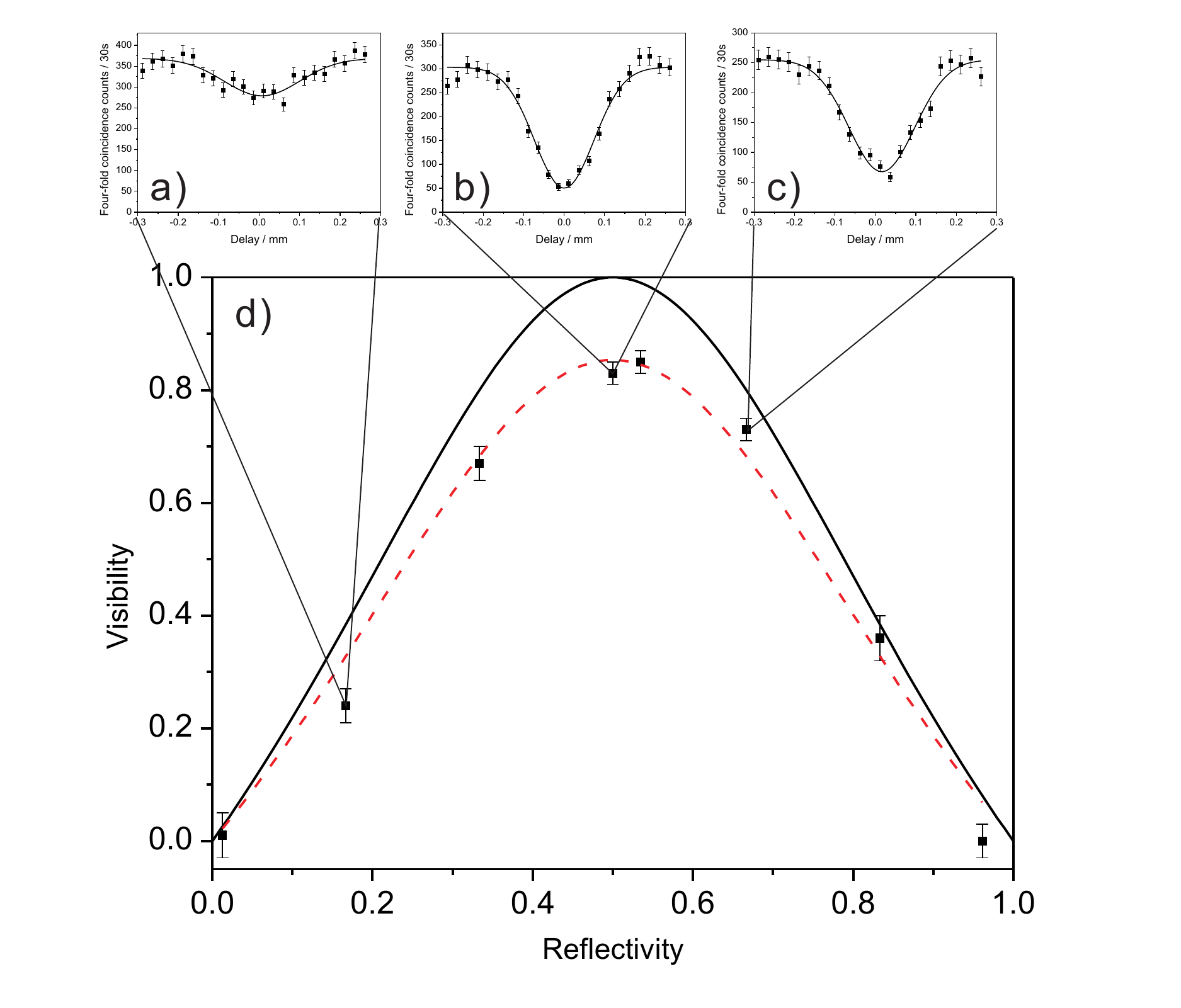}
\caption{\label{fig:TDCresults}
Experimental demonstration of variable measurement-induced interaction between two independent photons. (a), (b) and (c) Measured four-fold coincidence counts are plotted versus the relative optical delay between the interfering photons and fitted with a Gaussian function, when the reflectivities of the TDC are tuned to be 0.17, 0.5 and 0.67, respectively. (d) Visibility of the Hong-Ou-Mandel (HOM) dip when measuring four-fold coincidences using a TDC. The black squares are the experimental results. The black solid curve is a theoretical prediction based on Eq.\ \ref{visibility} and the red dashed curve is a fit with the only free parameter of $\textrm{V}_{sys}$, which is about 0.853. The main reason of this non-ideal value is due to the higher-order emission generated from SPDC sources. The error bars are based on a Poissonian statistics.}
 \end{center}
\end{figure}

Bunching due to the two-photon Hong-Ou-Mandel (HOM) interference~\cite{Hong1987} and the corresponding anti-bunching effect~\cite{Mattle1996} are crucial for many quantum information processing protocols, especially for photonic quantum computation experiments (C-phase gate~\cite{Kok2007,Pan2008}, entanglement swapping~\cite{Pan1998, Jennewein2002}, etc.), as well as for our photonic quantum simulation~\cite{Ma2011}. The bosonic nature of the photons shows up when the input states are superimposed at the TDC such that a detection, even in principle, cannot distinguish either of them. This leads to a superposition of double occupations on both outgoing modes and thus suppression of the coincidence detections, where one photon is detected in each output mode. The visibility of this HOM dip is one when the TDC is set to have equal splitting of transmitted and reflected photons similar to a 50/50 beam splitter. As soon as the two input photons can be partially distinguished by unbalancing the splitting ratio the visibility decreases. The dependence of the ideal visibility ($V_{ideal}$) upon the reflectivity of TDC ($\eta$) is the following:
\begin{equation}
\label{visibility}
V_{ideal}=\frac{2\eta(1-\eta)}{1 - 2 \eta + 2 \eta^2}.
\end{equation}
In Fig.\ \ref{fig:TDCresults} this reflectivity dependent ideal visibility is plotted in the black solid curve. Experimental imperfections due to high-order emissions from SPDC and group velocity mismatch reduce the visibility. Fig.\ \ref{fig:TDCresults} also shows the measured visibilities (black squares) and their corresponding fit (dashed red curve).

\section{Pair-wise entanglement dynamics}
The main advantage of the precise quantum control of individual particles is that inter-particle entanglement dynamics can be investigated. By using a similar experimental configuration as in Ref.\ \cite{Ma2011}, we study the entanglement distributions among different particles with respect to the effective interaction strength that was tuned by the TDC. For the quantification of the bipartite entanglement in our system, we use the measure of concurrence~\cite{Wootters1998}, which, for a given state $\rho$, is $C(\rho)\equiv
max\{0,\sqrt{\lambda_1} -\sqrt{\lambda_2}-\sqrt{\lambda_3} -
\sqrt{\lambda_4}\}$, where $\lambda_{i}$ are the eigenvalues of the matrix $\rho \Sigma \rho^{T}\Sigma$ in non-increasing order
by magnitude with $\Sigma=\sigma_{Y}\otimes \sigma_{Y}$, where
$\sigma_{Y}=-i\ket{0}\bra{1}+i\ket{1}\bra{0}$. While the previous quantum simulation characterized the pair-wise energy dynamics of Heisenberg interactions that were directly extractable from measured coincidence counts, this experiment requires the reconstruction of the density matrices to obtain the concurrence values. In the experiment we tune the reflectivity of the TDC and hence vary the photon-photon interaction strength. Various four-photon quantum states are tomographically measured and the density matrices of them are reconstructed~\cite{White1999,James2001}. The concurrence of the two-photon subsystems is calculated from the four-photon density matrices by tracing out the other two photons.

\begin{figure}
\includegraphics[width=0.9\textwidth]{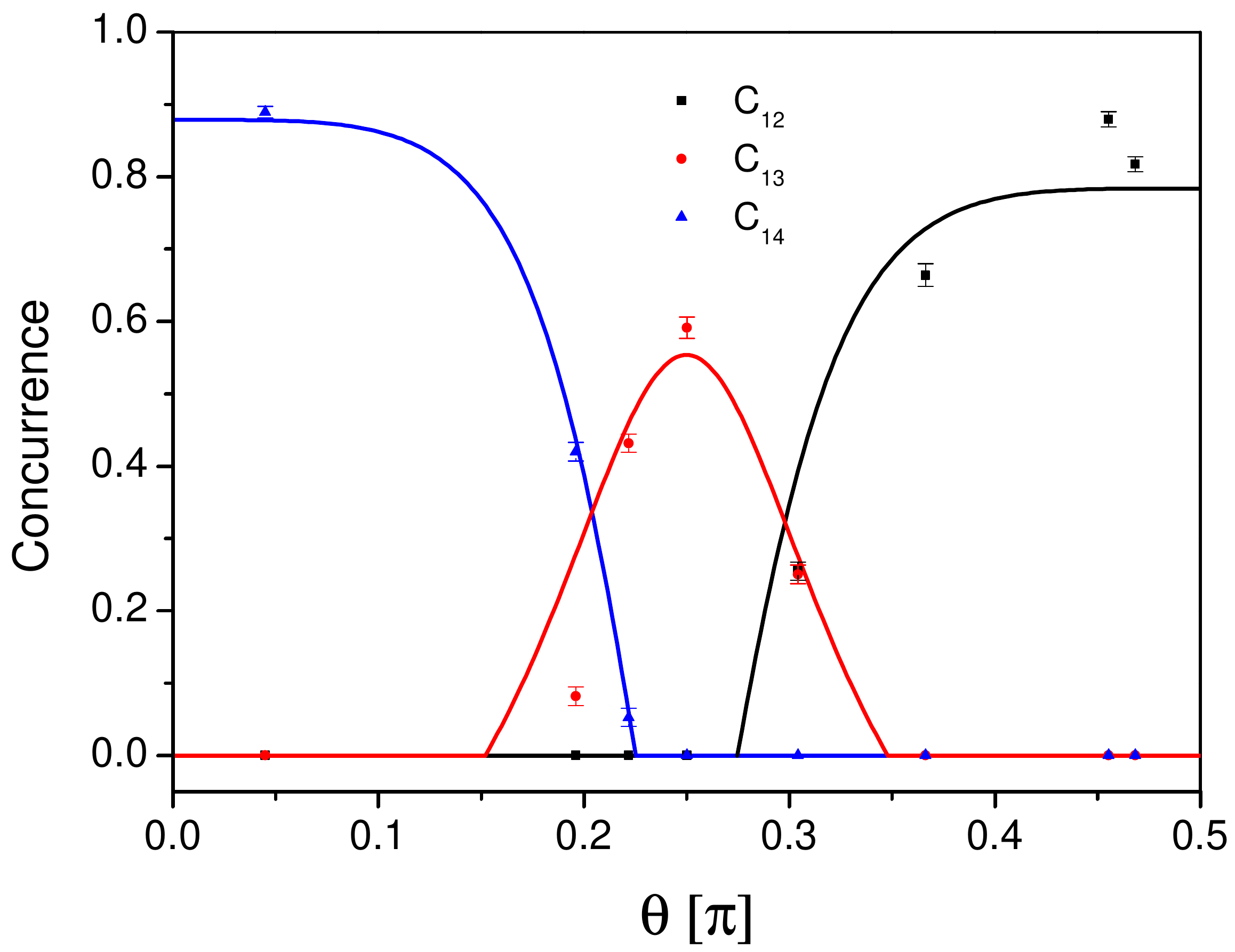}
\caption{\label{fig:Concurrence} Measured concurrence for various ground state configurations. Entanglement distribution between photons 1 \& 2 (black square, fitted with black curve), 1 \& 3 (red circle, fitted with red curve) and 1 \& 4
    (blue triangle, fitted with blue curve) are shown as a function of the TDC parameter $\theta$. The rise of entanglement in one pair causes the fall of the
    entanglement in other pairs according to monogamy. The monogamy gives rise of the entanglement sudden death and
    sudden birth \cite{Almeida2007, Yu2009, JimenezFarias2009}. We derive the uncertainties in concurrence from the density matrices, which are calculated using a Monte Carlo routine and assumed Poissonian errors.}
\end{figure}

Due to the quantum monogamy relations~\cite{Ma2011,Coffman2000a,Osborne2006} the total amount of pair-wise entanglement stays constant while the change of interaction strength affects the distribution and thus the ground state configurations. When tracking the change in entanglement by using concurrence, the expected ``sudden death" and ``sudden birth" of entanglement~\cite{Yu2009} can be seen (Fig.\ \ref{fig:Concurrence}). While this concept is typically used for studying environment-induced decoherence~\cite{JimenezFarias2009}, similar behavior can be observed here too. In fact our tunable interactions allow to mimic a controlled interaction with the environment of two additional particles, which opens the possibility to obtain insights into complex decoherence mechanisms. We explicitly show the concurrence of different photon-pair configuations with respect to the TDC angle $\theta$. The relation of the TDC angle and its reflectivity is given by $\theta=\arctan\sqrt{\eta}$. One can see that the concurrence for one photon pair, e.g. $C_{14}$ decreases rapidly as we increase $\theta$. Around $\theta=0.274$, all of a sudden the entanglement between another spin pair (e.g. $C_{13}$) is born at the cost of the reduced concurrence $C_{14}$, which vanishes as $\theta$ is further increased. The observation of similar disappearance or emergence of entanglement among the other photons demonstrates the capability of our quantum simulator for manipulating the quantum correlations~\cite{Almeida2007, Yu2009, JimenezFarias2009}.

\section{Generalized Heisenberg spin model on a four-site square lattice and a six-site checkerboard lattice}

The fact that the properties of photons seem to make the simulation of ground states for Heisenberg-interacting spins feasible, opens promising perspectives for the investigation of more complex interactions by using current technology. Therefore we investigate generalized Heisenberg models for the four-site square lattice and the six-site checkerboard lattice. For the four-site square lattice we extend the model in Ref.\ \cite{Ma2011} by adding a next-nearest neighbor interaction term (Fig 4a). The Hamiltonian for this system is:
\[
H=J_{1}(\mathbf{S}_{1}\cdot\mathbf{S}_{2}+\mathbf{S}_{3}\cdot\mathbf{S}_{4})+J_{2}(\mathbf{S}_{1}\cdot\mathbf{S}_{3}+\mathbf{S}_{2}\cdot\mathbf{S}_{4})+J_{3}(\mathbf{S}_{1}\cdot\mathbf{S}_{4}+\mathbf{S}_{2}\cdot\mathbf{S}_{3})
\]
\begin{figure}
 \begin{center}
\includegraphics[width=0.75\textwidth]{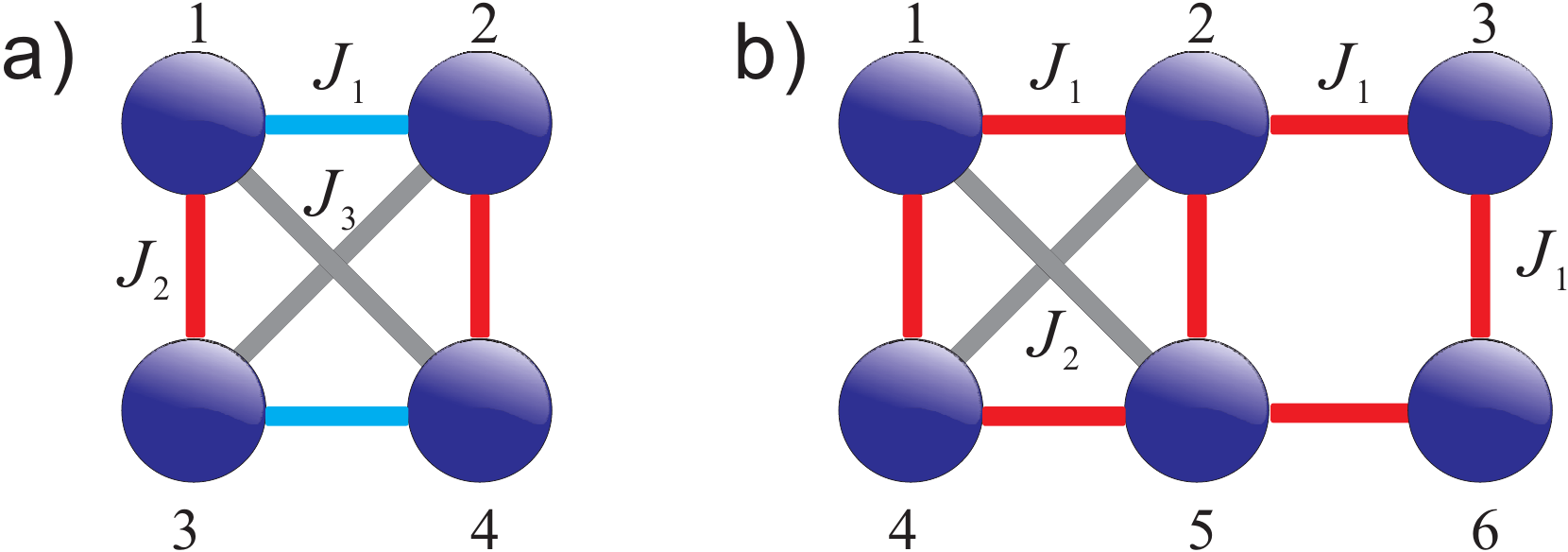}
\caption{\label{lattice4and6} a) A four-site square lattice with $J_{1}$ being horizontal coupling,
$J_{2}$ being vertical coupling and $J_{3}$ being diagonal coupling. b) A six-site checkerboard lattice. $J_{1}$ is the coupling between the nearest neighbors and $J_{2}$ is the coupling on the cross bonds. }
 \end{center}
\end{figure}

We consider the antiferromagnetic case with the couplings $J_{1}$, $J_{2}$, $J_{3}\ge0$.
For each of the three terms in $H$, the ground state is a pair of
singlets, $|\Phi_{=}\rangle=|\Psi_{12}^{-}\rangle|\Psi_{34}^{-}\rangle$,
$|\Phi_{||}\rangle=|\Psi_{13}^{-}\rangle|\Psi_{24}^{-}\rangle$ and
$|\Phi_{\times}\rangle=|\Psi_{14}^{-}\rangle|\Psi_{23}^{-}\rangle=|\Phi_{=}\rangle-|\Phi_{||}\rangle$.
In Ref. \cite{Ma2011} we showed that by tuning $J_{2}/J_{1}$ from
$0$ to $\infty$, the ground state, $\Phi_{g}$, gradually changes from $|\Phi_{=}\rangle$
to $|\Phi_{||}\rangle$.

\begin{figure}
 \begin{center}
\includegraphics[width=0.33\textwidth]{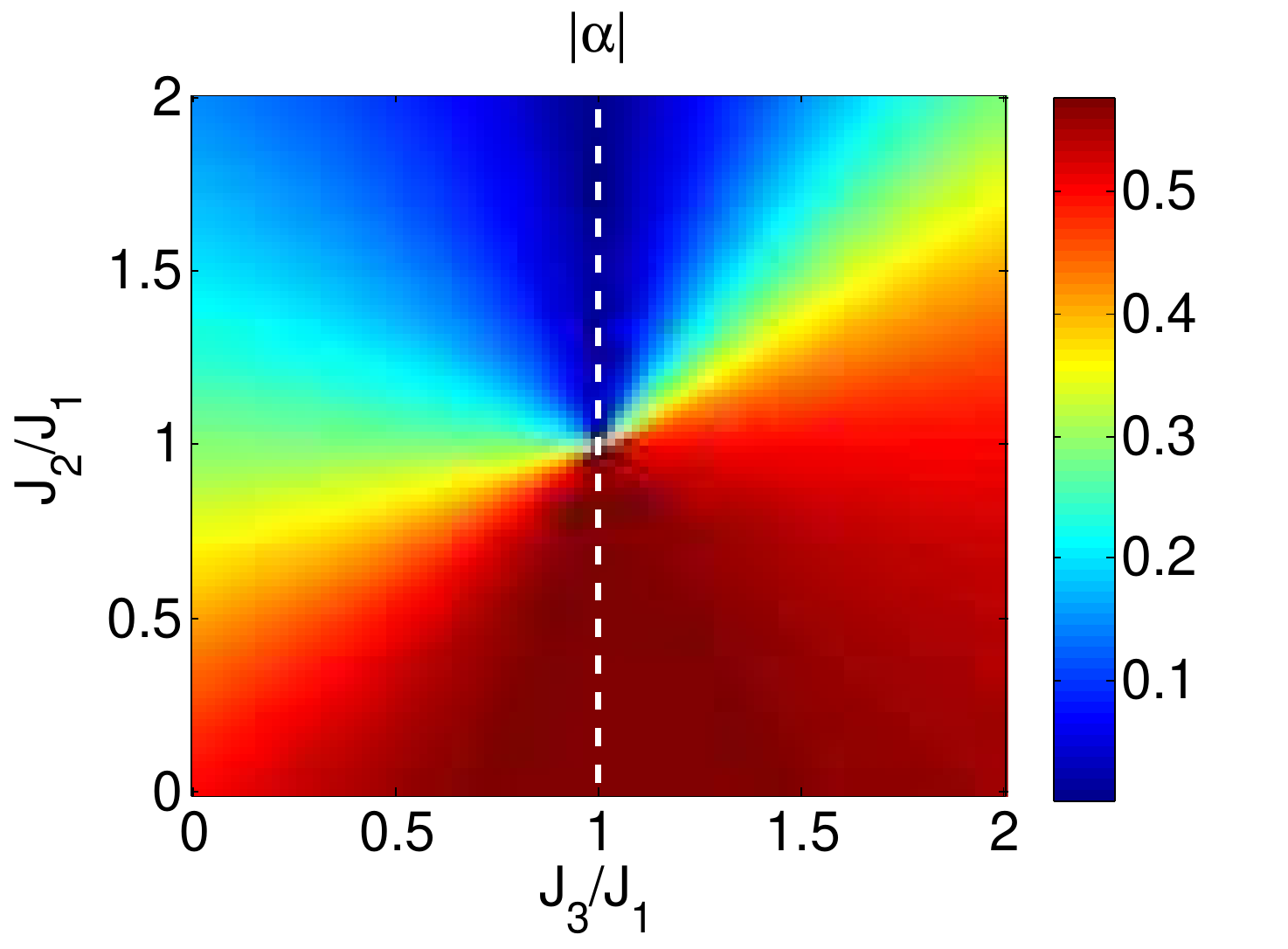}\includegraphics[width=0.33\textwidth]{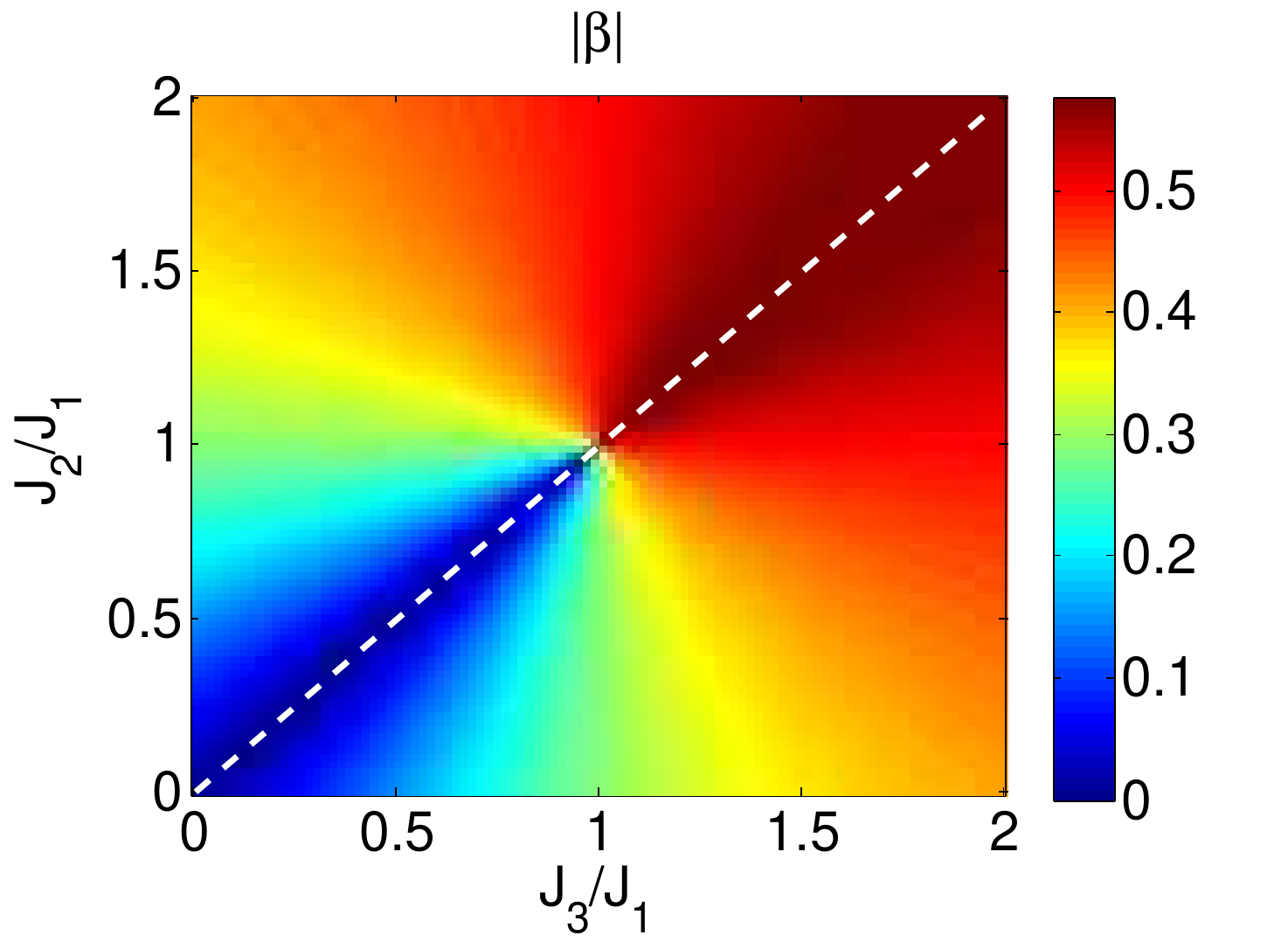}\includegraphics[width=0.33\textwidth]{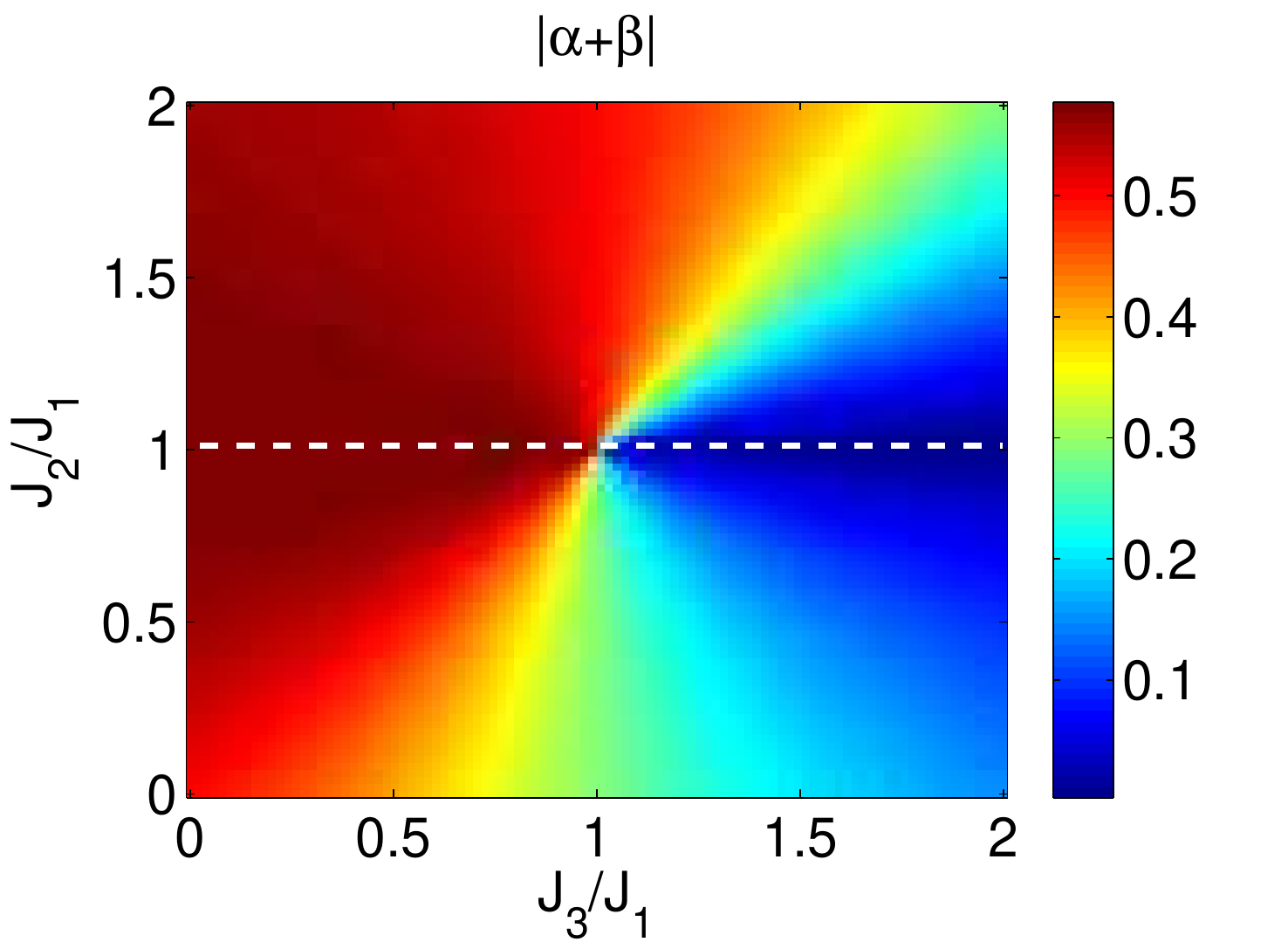}
\caption{\label{phase} Ground state phase diagram for various parameter regimes. The coefficients of the ground state, $|\Phi_{g}\rangle=\alpha|\Phi_{=}\rangle+\beta|\Phi_{||}\rangle$, are shown for different ratios of $J_{2}/J_{1}$ and $J_{3}/J_{1}$. In \textbf{a}), \textbf{b}) and \textbf{c}), we show $|\alpha|$, $|\beta|$ and $|\alpha|+|\beta|$, respectively. A clear phase transition can be seen along the dashed line \textbf{a}) for $J_{1}=J_{3}$, \textbf{b}) for $J_{2}=J_{3}$, and \textbf{c}) for $J_{1}=J_{2}$.}
 \end{center}
\end{figure}

With the introduction of $J_{3}$, the ground state of the system
is still a superposition $|\Phi_{g}\rangle=\alpha|\Phi_{=}\rangle+\beta|\Phi_{||}\rangle$ with normalization condition $2(|\alpha|^{2}+|\beta|^{2}+|\alpha+\beta|^{2})=1$.

Remarkably, tuning $J_{3}/J_{1}$ can induce sharp phase
transitions with a sudden change of the ground state configuration due to the competing of the valence bond configurations. In Fig.\ 5, the ground state configurations for different coupling regimes are shown. There are three particular interesting phase transitions:
\begin{itemize}
\item For $J_{1}=J_{3}$, $|\Phi_{g}\rangle$ suddenly changes from $|\Phi_{=}\rangle+|\Phi_{\times}\rangle$
to $|\Phi_{||}\rangle$ when $J_{2}/J_{1}$ is tuned across $1$
\item For $J_{2}=J_{3}$, $|\Phi_{g}\rangle$ suddenly changes from $|\Phi_{=}\rangle$
to $|\Phi_{||}\rangle-|\Phi_{\times}\rangle$ when $J_{2}/J_{1}$
is tuned across $1$
\item For $J_{1}=J_{2}$, $|\Phi_{g}\rangle$ suddenly changes from $|\Phi_{=}\rangle+|\Phi_{||}\rangle$
to $|\Phi_{\times}\rangle$ when $J_{3}/J_{1}$ is tuned across $1$
\end{itemize}

The case with $J_{1}=J_{2}\ne J_{3}$ is widely studied for square
lattice systems due to its relevance to cuprates, Fe-based superconductors,
and other materials~\cite{Fang08,Melzi08}. Previous studies have shown
that in the thermodynamic limit, when $J_{3}>J_{1}$, the system is in
a diagonal Neel ordered state~\cite{Jiang11}, which is consistent with the ground
state $|\Phi_{\times}\rangle$ for a minimum of four sites, as discussed
above. In the regime where $J_{2}<J_{1}$ the configuration for the ground state is still under debate due to regions that appear to be non magnetic. Numerical calculation have recently shown
that this region is highly likely to be a quantum spin liquid with $Z_{2}$ topological order~\cite{Jiang11}.
However, for only four spins, the region with $J_{2}<J_{1}$ has only a single ground state configuration $|\Phi_{=}\rangle+|\Phi_{||}\rangle$.

\begin{figure}[t]
 \begin{center}
\includegraphics[width=0.8\textwidth]{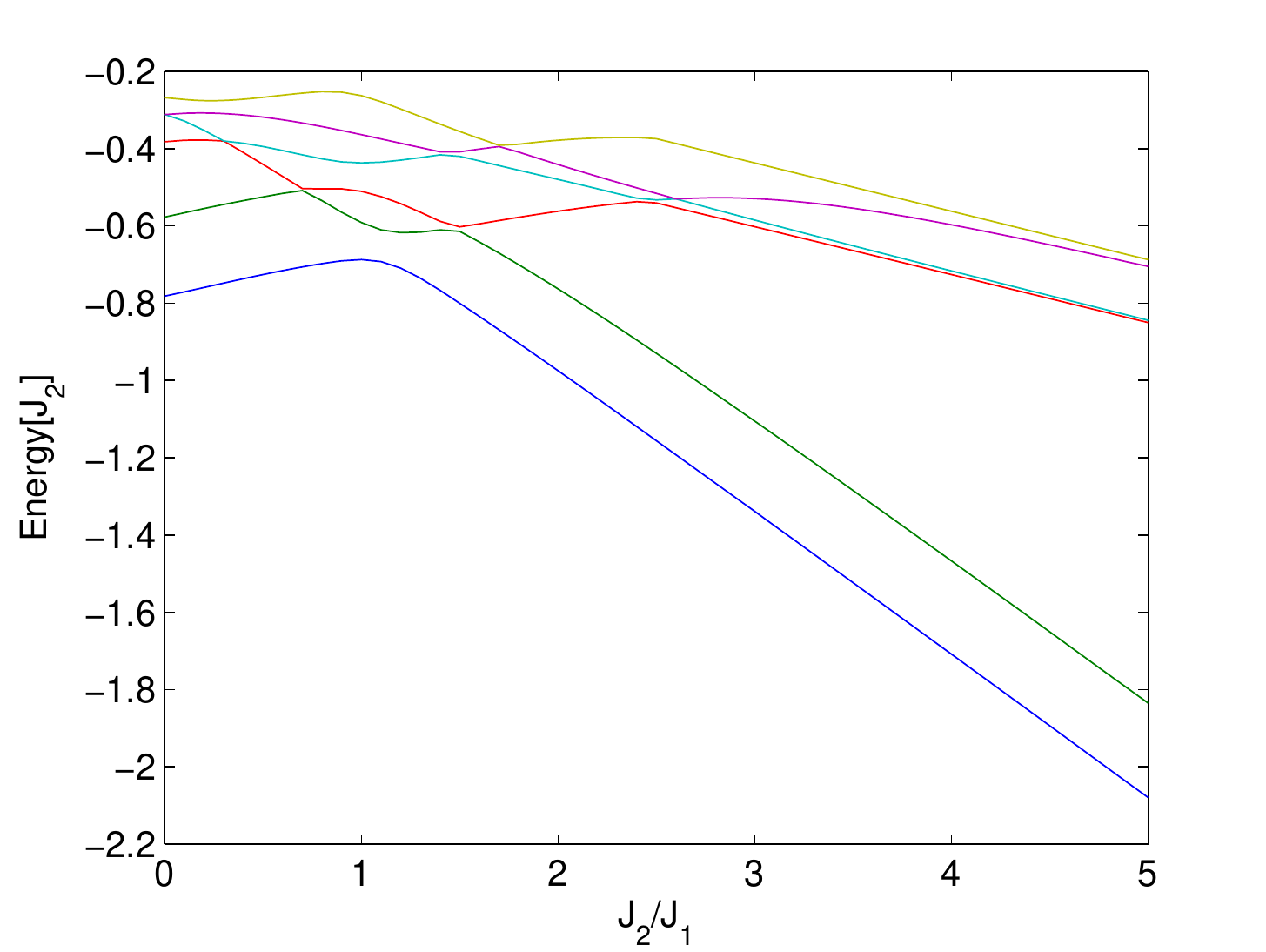}
\caption{\label{lattice6Energy} Energy spectrum of the low lying eigenstates of the six-site checkerboard lattice system. The energy levels are shown as a function of the ratio $J_{2}/J_{1}$. For the ratio $J_{2}/J_{1}=1$ an avoid level crossing of the ground state occurs.}
 \end{center}
\end{figure}

We also investigate theoretically the ground states of a $J_{1}-J_{2}$ Heisenberg model on a
six-site checkerboard lattice. The geometry of this system is shown
in Fig.\ \ref{lattice4and6}b, where $J_{1}$ is the coupling strength between the nearest
neighbor sites and $J_{2}$ is the coupling strength on the cross
bonds. The coupling ratio $J_{2}/J_{1}$ is the only parameter of
this system. The introduction of the next nearest neighbor coupling
$J_{2}$ makes this system a simple frustrated magnetic model. The
study of this model on thermodynamics limits is motivated by the three-dimensional
pyrochlore materials~\cite{Axtell97,Coldea01}. The two-dimensional model has been studied by
several groups~\cite{1,2,3,4,5,6,7}. It is known that for the regime where the coupling ratio $J_{2}/J_{1}\ll1$
the system has a co-linear Neel order. At $J_{2}/J_{1}\approx1$ numerical
calculation~\cite{2} suggests a plaquette valence bond solid ground
state while the ground state at $J_{2}/J_{1}\gg1$ remains under debate~\cite{1,3,4}. For this reason future photonic quantum simulations might provide answers to these open questions.

\begin{figure}[t]
 \begin{center}
\includegraphics[width=0.6\textwidth]{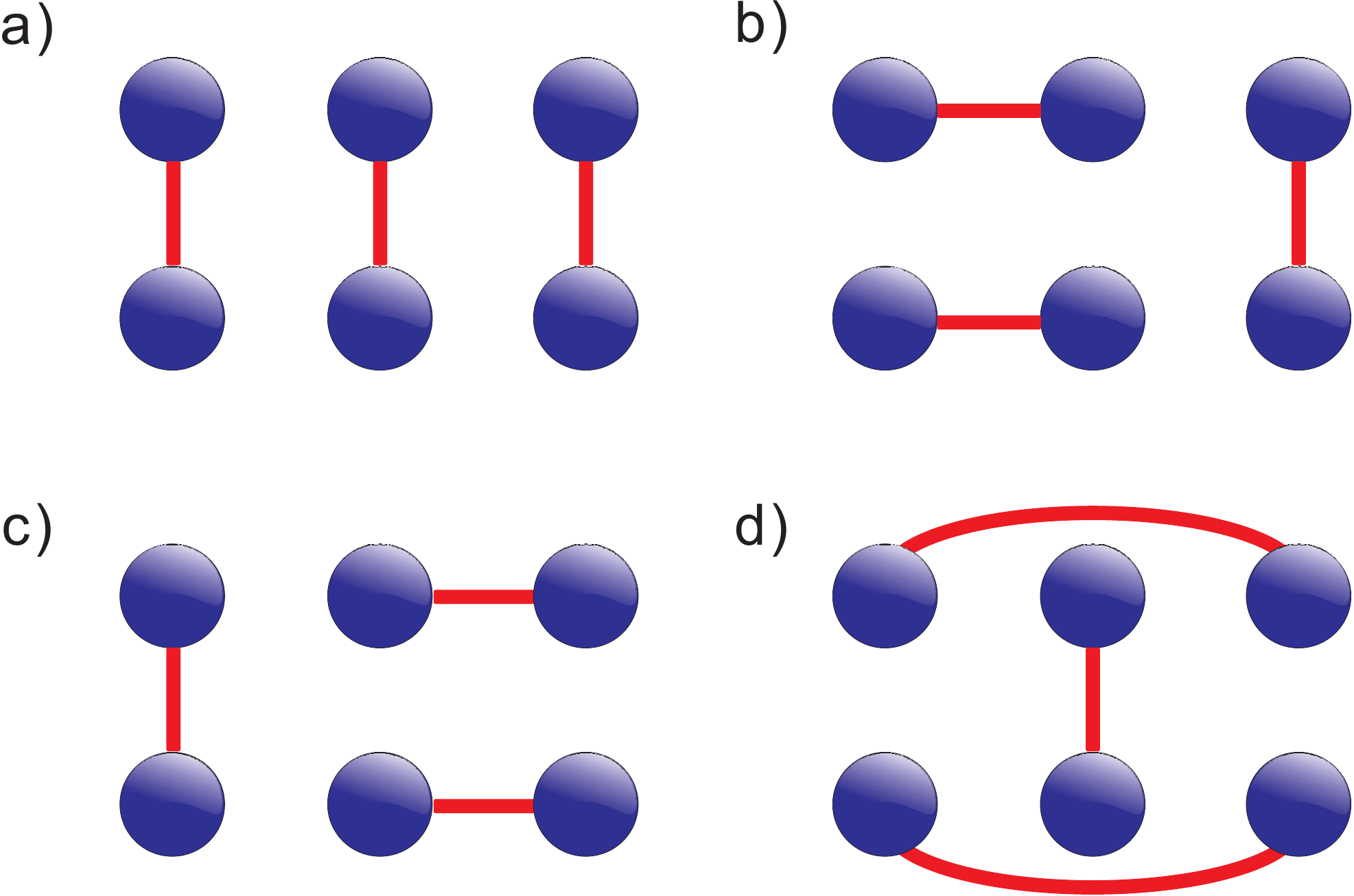}
\caption{\label{fourconfig} The independent ground state configurations of the six-site checkerboard lattice. The four  different dimer coverings are labeled as a)$\ket{\psi_{1}}$, b)$\ket{\psi_{2}}$,
c)$\ket{\psi_{3}}$, and d)$\ket{\psi_{4}}$ in our system.}
 \end{center}
\end{figure}

Numerical studies of the six-site checkerboard system were done by diagonalizing the Hamiltonian with open boundary conditions. In Fig.\ \ref{lattice6Energy} the energy spectrum of the six low lying states as a function of the ratio $J_{2}/J_{1}$ in the $\sum_{i}S_{i}^{z}=0$ subspace is presented. The energy spectrum shows an avoid level crossing around $J_{2}/J_{1}\approx1,$
which indicates a dramatic change of the ground state properties there. In analogy to the $J_{1}-J_{2}$ Heisenberg model on a plaquette, where the ground states can be expressed as linear superposition of two different dimer coverings whose coefficients depend on the ratio $J_{2}/J_{1}$, the ground states for the six-site checkerboard lattice can also be expressed as superpositions of various dimer configurations. For the discussed six-site lattice system, fifteen different dimer coverings exist. However, only six are independent. By taking the symmetry of our
system into consideration only four out of the six coverings are allowed (Fig.\ \ref{fourconfig}).

\begin{figure}[t]
 \begin{center}
\includegraphics[width=1\textwidth]{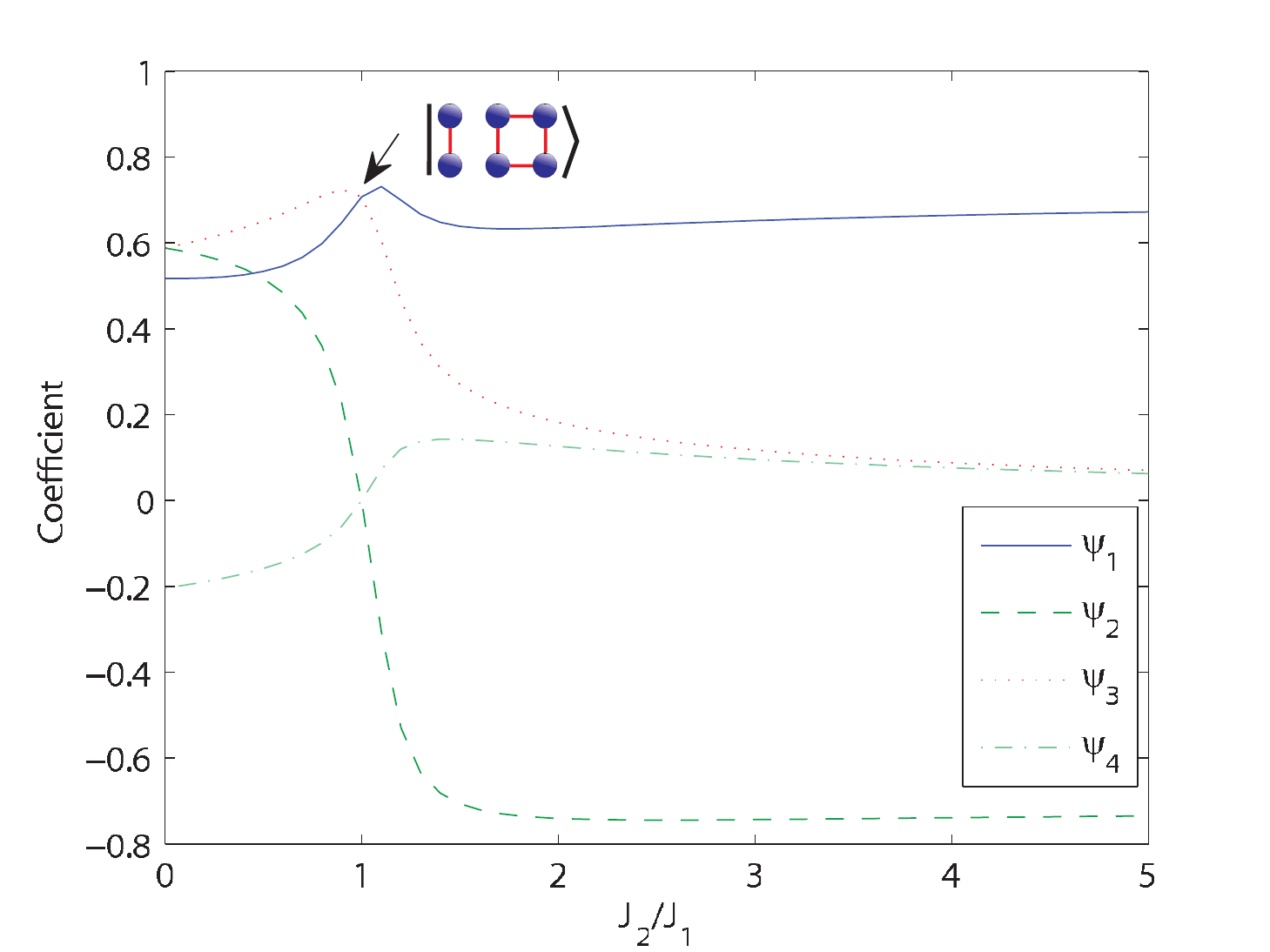} \caption{\label{Coefficients} The six-site checkerboard lattice ground state. The according superposition of the four independent dimer configurations, $\ket{\psi_{1}}$, $\ket{\psi_{2}}$, $\ket{\psi_{3}}$, $\ket{\psi_{4}}$ is shown as function of the ratio $J_{2}/J_{1}$. The inset shows the plaquette ground state at $J_{2}/J_{1}=1$.}
 \end{center}
\end{figure}

Therefore, the ground states can be described as superpositions
of four dimer coverings, $\ket{\psi_{1}}$, $\ket{\psi_{2}}$, $\ket{\psi_{3}}$, $\ket{\psi_{4}}$ in all region of $J_{2}/J_{1}$. In Fig.\ \ref{Coefficients} we show the contribution of each dimer configuration with respect to the coupling ratio $J_{2}/J_{1}$. At $J_{2}/J_{1}=1$ the coefficients
for $\ket{\psi_{1}}$ and $\ket{\psi_{3}}$ are equal and coefficients for $\ket{\psi_{2}}$
and $\ket{\psi_{4}}$ are exact zero. In our current convention of the dimer
covering wave function this particular superposition gives us a plaquette
state on the right four sites as it is shown in the inset. This coincides with the plaquette valence bond solid state in an
infinite system. Remarkably, the six-site checkerboard lattice thus provides already a valuable hint for the true ground state in the thermodynamic limit. The ground states at the ratio $J_{2}>J_{1}$ have close to
equal contribution from $\ket{\psi_{1}}$ and $\ket{\psi_{2}}$, which suggests
a significant contribution from a cross-dimer state on the left plaquette.
We would like to mention that a cross-dimer ground state has also been suggested
as a potential ground state for large $J_{2}$ couplings~\cite{3}. It is interesting that only a small contribution from the dimer configuration $\ket{\psi_{4}}$ can be found in all the possible cases.

\section{Summary and Outlook}
In conclusion, today's available photonic quantum technology is reaching the stage where significant advantages arise for the simulation of particular interesting questions in solid-state physics and quantum chemistry. Therefore photonic quantum simulations provide exciting opportunities to cover, for example, the direct construction of custom-tailored many-body wave functions. Impressively, the usage of optical elements such as tunable non-polarizing or polarizing beam splitters enables entangled few-photon states to construct many-body valence bond wave functions for molecular and solid-state systems due to non-classical interferences. As we have shown above, this is of particular interest in condensed matter physics as it provides insight into the frustration of strongly-correlated spin systems and the onset of quantum phase transitions. On the other hand, in quantum chemistry it effectively allows studying delocalized bonds in chemical structures and chemical reactions~\cite{Aspuru-Guzik2005}. Thus being able to monitor the full dynamics of individual particles and bonds provide some fascinating perspective for the quantum simulation of small molecules or reactive centers.

The main future challenge will be to increase the number of photons or degrees of freedom to realize a sufficient amount of qubits such that quantum computers can outperform their classical counter parts. In general, up to the level of approximately twenty qubits it presently appears possible to conceive a system based on bare physical qubits. However, given the current experimental limitations, operational fidelities and noise sources, it seems that useful system consisting of more than twenty qubits could not be realized without some level of error correction. But in contrast to the implementations of well-known quantum algorithms, such as Shor's algorithm for a computationally relevant key-length, the requirements for fault-tolerance are much less demanding. How much error correction is needed to achieve a useful quantum simulation is an open problem.

Recent work~\cite{Aspuru-Guzik2005,Lanyon2010} has shown that quantum systems with less than a dozen physical qubits are capable of simulating chemical systems with a precision that cannot be achieved by conventional computers, when processed via almost a thousand discrete gate operations. Although such small quantum systems are feasible by using present quantum technology, the requirements in terms of gate operations is tremendous. Thus, analog quantum gate operations look promising in reducing the technical complexity of performing such quantum simulation experiments by requiring fewer number of physical gates.

\ack
XSM, BD, SK, WN, YHC, AZ and PW thank F. Verstraete and \v{C}. Brukner for helpful discussions. We acknowledge support from the European Commission, Q-ESSENCE (No. 248095), ERC Advanced Senior Grant (QIT4QAD) and the ERA-Net CHIST-ERA project QUASAR, the John Templeton Foundation, the Austrian Nano-initiative NAP Platon, the Austrian Science Fund (FWF): [SFB-FOCUS] and [Y585-N20] and the doctoral programme CoQuS, and from the Air Force Office of Scientific Research, Air Force Material Command, USAF, under grant number FA8655-11-1-3004. YHC, ZXG, and LMD thank H.~-C. Jiang for helpful discussion and acknowledge support from the NBRPC (973 Program) 2011CBA00300 (2011CBA00302)  and the DARPA OLE program.

\section*{References}


\end{document}